\documentclass[10pt, final, journal, twocolumn]{IEEEtran}
\usepackage{times}
\usepackage{amsmath}
\usepackage{amssymb}
\usepackage{verbatim}
\usepackage{enumerate}
\usepackage{graphicx}
\usepackage{algorithm}
\usepackage{algpseudocode}
\usepackage{cite}
\usepackage{hyperref}
\usepackage{mathtools}

\usepackage{xcolor}
\usepackage{listings}
\definecolor{codegreen}{rgb}{0,0.6,0}
\definecolor{codegray}{rgb}{0.5,0.5,0.5}
\definecolor{codepurple}{rgb}{0.58,0,0.82}
\definecolor{backcolour}{rgb}{0.95,0.95,0.92}

\lstdefinestyle{mystyle}{
  backgroundcolor=\color{backcolour},   commentstyle=\color{codegreen},
  keywordstyle=\color{magenta},
  numberstyle=\tiny\color{codegray},
  stringstyle=\color{codepurple},
  basicstyle=\ttfamily\footnotesize,
  breakatwhitespace=false,         
  breaklines=true,                 
  captionpos=b,                    
  keepspaces=true,                 
  numbers=left,                    
  numbersep=5pt,                  
  showspaces=false,                
  showstringspaces=false,
  showtabs=false,
  escapechar=|,
  frame=single,                  
  tabsize=2,
  xleftmargin=1.5em,
  framexleftmargin=1em
}

\lstdefinestyle{mystyle2}{
  backgroundcolor=\color{backcolour},   commentstyle=\color{codegreen},
  keywordstyle=\color{magenta},
  numberstyle=\tiny\color{codegray},
  stringstyle=\color{codepurple},
  basicstyle=\ttfamily\footnotesize,
  breakatwhitespace=false,         
  breaklines=true,                 
  captionpos=b,                    
  keepspaces=true,                                  
  numbersep=5pt,                  
  showspaces=false,                
  showstringspaces=false,
  showtabs=false,
  escapechar=|,
  frame=single,                  
  tabsize=2,
}

\ifCLASSOPTIONcompsoc
  \usepackage[nocompress]{cite}
\else
  \usepackage{cite}
\fi

\begin{document}
\title{An eBPF-Based Trace-Driven Emulation Method for Satellite Networks}
\author{Weibiao~Tian,~Ye~Li,~\IEEEmembership{Member,~IEEE},~Jinwei~Zhao,~Sheng~Wu,~\IEEEmembership{Member,~IEEE},~Jianping~Pan,~\IEEEmembership{Fellow,~IEEE}
\thanks{Weibiao Tian (jiuzhaotwb@stmail.ntu.edu.cn) is with School of Information Science and Technology, Nantong University, Nantong 226019, China.}
\thanks{Ye Li (yeli@ntu.edu.cn) is with School of Information Science and Technology, Nantong University, Nantong 226019, China, and also with Nantong Research Institute for Advanced Communication Technologies (NRIACT), Nantong 226019, China.}
\thanks{Jinwei Zhao (clarkzjw@uvic.ca) and Jianping Pan (pan@uvic.ca) are with Faculty of Engineering and Computer Science, University of Victoria, Victoria BC V8P 5C2, Canada.}
\thanks{Sheng Wu (thuraya@bupt.edu.cn) is with the School of Information and Communication Engineering, Beijing University of Posts and Telecommunications, Beijing 100876, China.}
\thanks{\textit{Corresponding Author: Ye Li}}
}

\maketitle

\begin{abstract}
System-level performance evaluation over satellite networks often requires a simulated or emulated environment for reproducibility and low cost. However, the existing tools may not meet the needs for scenarios such as the low-earth orbit (LEO) satellite networks. To address the problem, this paper proposes and implements a trace-driven emulation method based on Linux's eBPF technology. Building a Starlink traces collection system, we demonstrate that the method can effectively and efficiently emulate the connection conditions, and therefore provides a means for evaluating applications on local hosts.
\end{abstract}

\setcounter{page}{1}
\thispagestyle{headings}

\section{Introduction}
System-level performance evaluation is a key part in network research, where a network is simulated to create end-to-end connections with reproducible conditions to readily evaluate applications. 
Using discrete-event network simulators such as ns-3, OMNET++ is popular, where connections over a virtual network with virtual protocol stacks can be created. However, one limitation of this approach is that a protocol module has to be implemented first. While the existing network simulators have supported various protocol modules from computer and mobile networks to more specialized ones (e.g., underwater acoustic networks), the development of the modules for satellite networks has been considerably lagging behind the network construction. For example, the Starlink has been in operation since 2021, and the integration with low-earth-orbit (LEO) networks is also a hot topic in the 6G evolution under 3GPP's term of non-terrestrial networks (NTNs), but there are very few Starlink or NTN modules publicly available for the main-stream network simulators so far. While there is a third-party ns-3 module \texttt{sns3} for the geostationary scenario with DVB-S2/DVB-RCS2 support \cite{Puttonen2014SimuTools}, most of the existing tools for LEO only simulate the topology and mobility \cite{Kassing2020IMC}, \cite{Kempton2021ICC}, without underlying protocol.

Network emulation (e.g., Mininet, Dummynet) is another approach, where a virtual topology is created, but with real TCP/IP stacks over Ethernet and the packets go through the Linux kernel. Queue disciplines such as \texttt{netem} are then employed at the traffic control (TC) subsystem to directly emulate connection conditions, without actually implementing the underlying protocol procedures of a specific network. This approach is advantageous in that the applications and the communications are real-time, and hence is capable of evaluating real-life performance. However, one limitation is that the emulators require statistical models to specify the random behavior of packet delay and losses. Unfortunately, the delay/loss behavior of the satellite networks have been shown to be quite unique. A typical example is the LEO scenario, where the handover of satellites causes highly dynamic connection-level delay and losses \cite{Ma2023Infocom}, \cite{Pan2024ICC}, for which the existing models in \texttt{netem} is far from sufficient to characterize. 

Given the aforementioned limitation, evaluating applications over a satellite network has been difficult given that the access to a physical satellite network (e.g., Starlink) is limited or too expensive for the majority of researchers. Devising an alternative method to effectively evaluate the network at low cost is highly desirable. To meet this demand, this paper proposes and implements a simple yet very effective evaluation method, which is driven by end-to-end traces recording the delay/loss of each packet. The trace may be collected from real networks using measurement tools and/or synthesized by artificial intelligent (AI) generative models trained from large trace datasets. For efficient emulation, the eBPF (extended Berkley Packet Filter) technology \cite{Vieira2020ACMSurvey} of Linux kernel is exploited to replay the per-packet trace in real time. As a demo use case, we build a Starlink testbed and present an emulation example. We demonstrate that the approach can effectively reproduce the connection conditions of LEO satellite networks on local hosts. To the best of our knowledge, there exist no publicly available alike tools for such emulation needs. We believe that the method may significantly facilitate application evaluation over satellite networks for the community. 

\section{System Design}
\subsection{Problem Statement}
Our goal is to emulate packet-level delay/loss conditions according to a real-life trace on a local host/network. Specifically, suppose that we are given a trace of an end-to-end connection,\footnote{The connection may refer to a packet flow of either connection-oriented (e.g., TCP, QUIC) or connection-less protocols (e.g., UDP).} which records the detailed delay and losses of the packets $\mathbf{p}_1,\mathbf{p}_2,\ldots$ sent over a network and received on the other end. For ease of description, assume that the trace, after proper data processing, is in the form of $\mathcal{T}=\{t_1, t_2,\ldots\}$, where $t_i>0$ is the delay between sending and receiving $\mathbf{p}_i$, and $t_i=-1$ if the packet was lost. The task is to replay the trace over a peer-to-peer link, either between two virtual machines on the same host or between physical hosts on the same local-area network with negligible propagation delay.

\subsection{eBPF-Based Emulation Architecture}
eBPF is a technology of Linux that allows to run programs in the operating system's kernel space at runtime, without requiring changes to the kernel source code or loading modules. eBPF programs are attached to several pre-defined hooks exposed by the kernel, thereby reading/modifying kernel-level data to extend the kernel's capabilities. The eBPF programs are verified before loading and support just-in-time compilation of its bytecode to native instructions, so the programs are secure and fast. eBPF has been a key enabler of many new applications in networking, monitoring, and security since its introduction in 2014. Via eBPF maps, which are specific data structures of efficient key/value stores, data can be efficiently shared between the user and kernel spaces.

\begin{figure}[hbt]
  \includegraphics[width=3.4in]{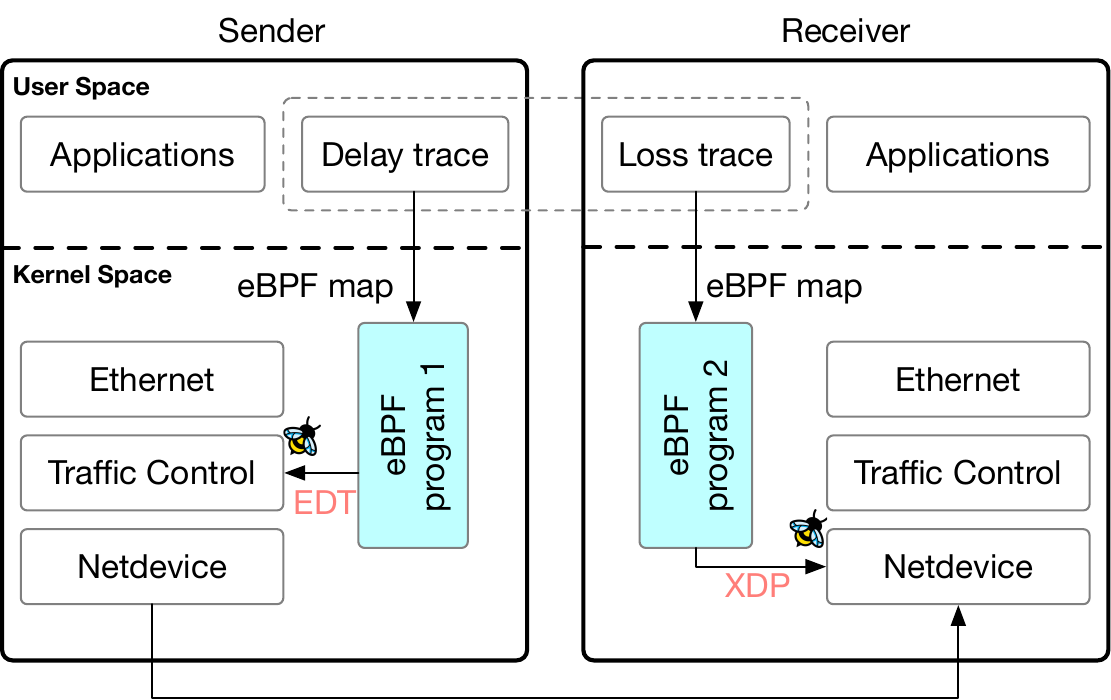}
  \caption{Architecture of the eBPF-based emulation}\label{fig_architecture}
\end{figure}

Using eBPF, we propose the architecture in Fig. \ref{fig_architecture} to achieve the emulation goal, where the delay and losses are emulated at the outbound (\texttt{egress}) and inbound (\texttt{ingress}) points of the network devices of the sending and receiving hosts, respectively. Given trace $\mathcal{T}$, we can obtain two separate traces, namely the delay trace $\mathcal{T}_d=\{t_1^\prime,t_2^\prime,\ldots\}$ and the loss trace $\mathcal{T}_l=\{I_1,I_2,\ldots\}$, defined as
\begin{equation*}
	t_i^\prime =\begin{cases}
		t_i & t_i>0,\\
		t_{i-1} & t_i=-1;
	\end{cases}\quad\quad
	I_i=\begin{cases}
		0 & t_i>0,\\
		1 & t_i=-1.
	\end{cases}
\end{equation*}
Note that a lost packet does not have a valid delay recorded in the original trace. To emulate the packet transmission and the loss, here we designate the delay of a lost packet the same as that of its previous packet. This ensures that the packet to be dropped arrives at the receiver in the original order.

 To emulate $\mathcal{T}_d$, an eBPF program is attached to the kernel's traffic control (TC) subsystem on the sending host, acting as the egress classifiers. Using the EDT (Earliest Departure Time) model \cite{Saeed2017Sigcomm}, delay can be emulated by setting the corresponding outbound packet's departure time \cite{Becker2022IC2E}. The packet is then scheduled to send by a timing wheel in the TC subsystem.
 
 To emulate $\mathcal{T}_l$, another eBPF program is devised to attach to the XDP (eXpress Data Path) hook of the kernel's RX path on the receiving host. The XDP is in the NIC (network interface controller) driver. It is right after the packet is received from the wire and is before any stack-related kernel memory allocation. By reading the loss trace via the eBPF map, the corresponding lost packets are dropped. Thanks to the XDP's design, these can be achieved at a hardware rate of more than millions of packets per second.

\section{Implementation}
This section highlights several key points of the method.\footnote{Upon publication of the paper, the full codes and the deployment procedures will be open-sourced at: \url{https://github.com/yeliqseu/ebpf-trace-emu}. During the peer-review, the materials are available upon requests.}

\subsection{Delay Emulation}
We first need to declare an eBPF map, as shown in Listing \ref{lst_ebpf_map}, to allow an eBPF program read traces from the kernel space at runtime. eBPF map does not allow dynamic memory allocation. That means the number of packets in the trace needs to be specified in the declaration (i.e., \texttt{TRACE\_LEN}). The map can be filled/updated after loading the program into the kernel, as will be described shortly.
\begin{lstlisting}[language=c, caption=eBPF map for delay trace, label=lst_ebpf_map, style=mystyle]
struct {
    __uint(type, BPF_MAP_TYPE_ARRAY);
    __uint(max_entries, TRACE_LEN);
    __type(key, __u32);
    __type(value, __u32);
} delay_map SEC(".maps");
\end{lstlisting}

Using the packet index as the key of the map, the essential codes of the eBPF program for delaying packets are shown in Listing \ref{lst_delay_packet}, where line \ref{lst_set_sec} names the program as \texttt{delay\_ebpf}, which will be used when attaching it to a hook. Each packet, contained in the kernel's \texttt{\_\_sk\_buff} structure, is first filtered according to the protocol and ports depending on the emulation needs. This is needed when one desires to only emulate delay for packets belonging to a certain TCP connection with a given source and destination port, so as to not affect other system traffic. To save space, this part is omitted. The core of the program is line \ref{lst_delay_start} to \ref{lst_delay_end}, where the corresponding delay of the packet is looked up from the map, and the delay is imposed by setting \texttt{tstamp} of the packet.

\begin{lstlisting}[language=c, caption={eBPF program (snippet) for delay, \texttt{edt\_delay\_packet.c}}, label=lst_delay_packet, style=mystyle]
static __u32 packet_index = 0;

SEC("delay_ebpf")|\label{lst_set_sec}|
int edt_delay_packet(struct __sk_buff *skb) {
  struct ethhdr eth;
  struct iphdr ip;
  
  /*|\label{lst_filter_start}|
   * Match skb according to protocol/ports
   */|\label{lst_filter_end}|
  
  // For a matched packet
	__u32 key = packet_index % TRACE_LEN;
	__u32 *delay_ns;
	packet_index++;
	if (packet_index >= TRACE_LEN) {
	  packet_index = 0;
	}
	delay_ns = bpf_map_lookup_elem(&delay_map, &key);|\label{lst_delay_start}|
	if (delay_ns) {
	  __u64 now = bpf_ktime_get_ns();
	  skb->tstamp = now + ((__u64) * delay_ns);
	}|\label{lst_delay_end}|
  // End of processing	
  return TC_ACT_OK;
}
\end{lstlisting}

\subsection{Loss Emulation}
The declaration of the eBPF map for the loss trace is similar to Listing \ref{lst_ebpf_map} and is omitted. The eBPF program for the XDP is shown in Listing \ref{lst_drop_packet}, where line \ref{lst_set_sec_xdp} indicates the hook point. Note that XDP is before \texttt{\_\_sk\_buff} allocation, and the packet is exposed in \texttt{xdp\_md} at this point. The procedure for processing and emulating the loss is similar to that for the delay. A packet is dropped if the eBPF program returns \texttt{XDP\_DROP}.

\begin{lstlisting}[language=c, caption={eBPF program (snippet) for loss, \texttt{xdp\_drop\_packet.c}}, label=lst_drop_packet, style=mystyle]
static __u32 packet_index = 0;

SEC("xdp_port_filter")|\label{lst_set_sec_xdp}|
int xdp_drop_packet(struct xdp_md *ctx) {
  void *data_end = (void *)(long)ctx->data_end;
  void *data = (void *)(long)ctx->data;
  struct ethhdr *eth = data;
    
  /*
   * Match ctx according to protocol/ports
   */
   
  // For a matched packet
  __u32 key = packet_index % TRACE_LEN;
  int *loss_flag;
  loss_flag = bpf_map_lookup_elem(&loss_map, &key);
  if (!loss_flag) {
    return XDP_PASS;
  }
  packet_index++;
  if (packet_index >= TRACE_LEN){
    packet_index = 0;
  }
  if (*loss_flag == 1) {
    return XDP_DROP;
  } else {
    return XDP_PASS;
  }
  // End of processing
}
\end{lstlisting}

\subsection{Deployment}
The deployment of the emulation follows a standard procedure of eBPF programs, which includes the compilation and attaching to the targeted hook. During the compilation, the compiler automatically invokes the eBPF verifier to perform static code analysis to ensure safety for the kernel: 
\begin{lstlisting}[language=bash, style=mystyle2]
clang -O2 -g -target bpf -c edt_delay_packet.c -o edt_delay_packet.o 
\end{lstlisting}

To attach \texttt{delay\_ebpf}, it requires to first add a dummy \texttt{clsact} qdisc to the targeted network device. Then, eBPF is set as the TC filter to directly act on the egress packets, with the program being attached. The commands are as follows:
\begin{lstlisting}[style=mystyle2]
sudo tc qdisc add dev enX1 clsact
sudo tc filter add dev enX1 egress bpf direct-action obj edt_delay_packet.o sec delay_ebpf
\end{lstlisting}

To realize the departure time specified by \texttt{delay\_ebpf}, the \texttt{fq} qdisc needs to be added as the root qdisc of TC, which finally turns on the EDT model:
\begin{lstlisting}[language=bash, style=mystyle2]
sudo tc qdisc add dev enX1 root fq
\end{lstlisting}

Compared to the delay program, attaching \texttt{xdp\_bpf} to XDP is relatively simpler by using \texttt{ip} from Linux's \texttt{iproute2} utilities after compilation, as shown below:
\begin{lstlisting}[language=bash, style=mystyle2]
clang -O2 -g -target bpf -c xdp_drop_packet.c -o xdp_drop_packet.o
sudo ip link set dev enX1 xdpgeneric obj xdp_drop_packet.o sec loss_bpf
\end{lstlisting}

The last remaining piece is to load the trace content to the eBPF maps. Various methods can be used, and \texttt{bpftool} is one convenient way to achieve this from outside the eBPF program. After the program is loaded, the corresponding ID of the declared map in the system can be identified using \texttt{bpftool map show} with the map name (e.g., \texttt{delay\_map} in Listing \ref{lst_ebpf_map}). With this ID, the following command can conveniently update the map using data from the trace files.

\begin{lstlisting}[language=bash, style=mystyle2]
sudo bpftool map update id {map_id} key {key_list} value {value_list}
\end{lstlisting}

To accurately emulate the trace, we remark that extra care should be taken for packet reordering, which would occur when the delay is highly dynamic. Since we have used packet index as the key for both delay and loss emulation, the loss trace needs to be properly re-ordered according to the arrival order of the packets before loading into the map. This can be easily done according to the delay trace. All the above commands can be wrapped up in one script to automate the deployment. The scripting details are omitted.

\section{Emulating Starlink Connection}
\subsection{Trace Collection}
\begin{figure}[hbt]
  \includegraphics[width=3.4in]{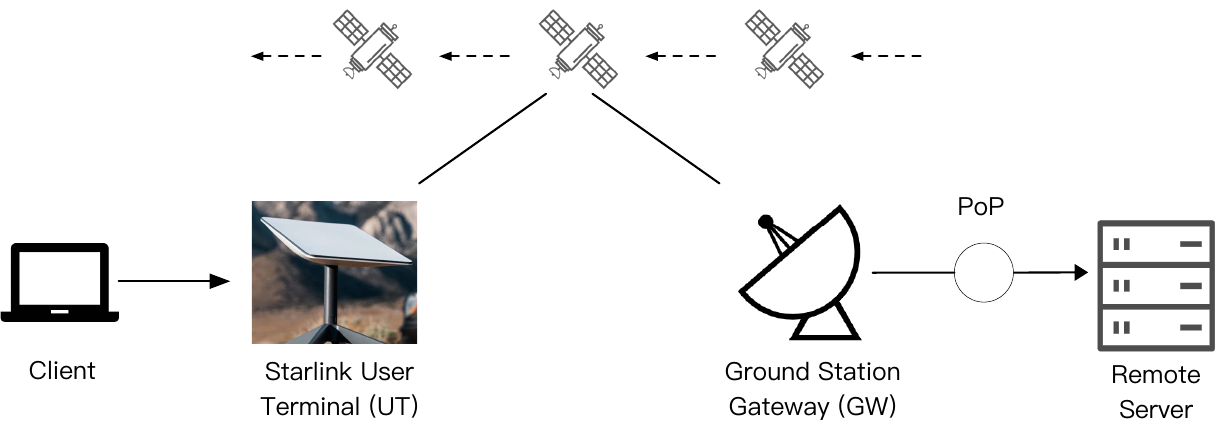}
  \caption{Starlink trace collection using \texttt{irtt}}\label{fig_irtt_setup}
\end{figure}
We setup a trace collection system as shown in Fig. \ref{fig_irtt_setup} using \texttt{irtt} \cite{irtt}, where a client, wired to a Starlink user terminal (UT) via a local user router, accesses a server on a remote host close to a Starlink PoP (point of presence). The UT is installed at Victoria, BC, Canada. To reduce the interference from the wild Internet, the server is set up colocated to a PoP that is closest to Victoria, which is in Seattle, USA. With this settings, the end-to-end Starlink connection traverses one hop bent pipe link and does not include any inter-satellite links (satellite handover still occurs). The client sends $60$-bytes packets every $10$ms and collects echo from the server, obtaining detailed traces of delay and loss for each packet in the \texttt{json} format. The traces can be readily processed to other desired form.

\subsection{Emulation and Verification}
We pick a $12$-hour trace collected from 00:00AM-12:00PM on August 23, 2024.\footnote{Our monthly-updated Starlink trace dataset can be accessed at \url{https://github.com/clarkzjw/LENS}.} With \texttt{irtt}, we obtain the round-trip time (RTT) and the packet loss traces. 

To perform the trace-driven emulation, we set up two virtual machines (VMs) on a private cloud, referred to as nodes A and B, respectively. The native RTT between the VMs is less than $1$ms and is hence negligible.\footnote{Using tools such as Mininet can also do the task, where VMs are emulated on the same physical machine via process-based virtualization.} To emulate the connection, we run the eBPF delay and the loss programs on node A and B, respectively. To verify, we then run the \texttt{irtt} client and server on A and B, respectively, to collect the delay/loss traces over the emulated connection.

\begin{figure}[hbt]
  \includegraphics[width=3.4in]{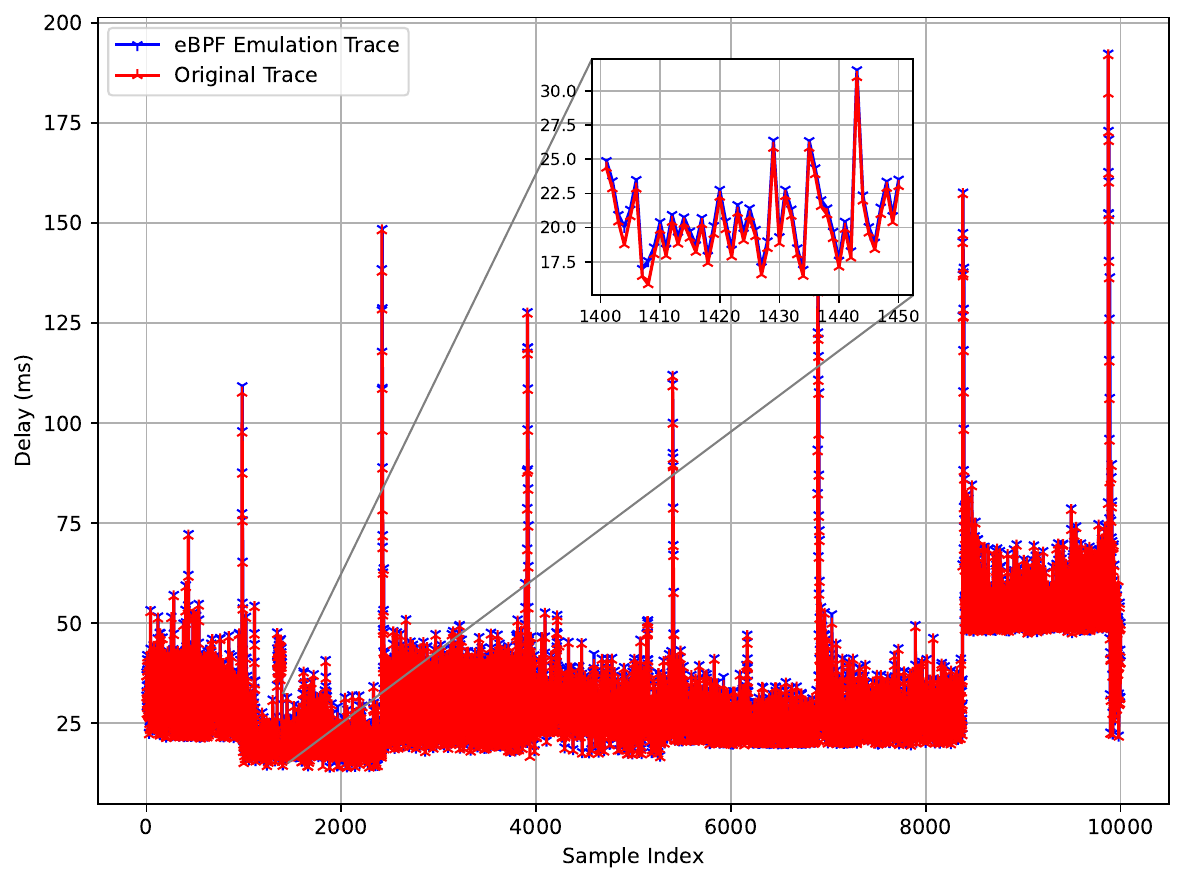}
  \caption{Comparison of the emulated and original delay traces}\label{fig_delay_comparison}
\end{figure}

\begin{figure}[hbt]
  \includegraphics[width=3.4in]{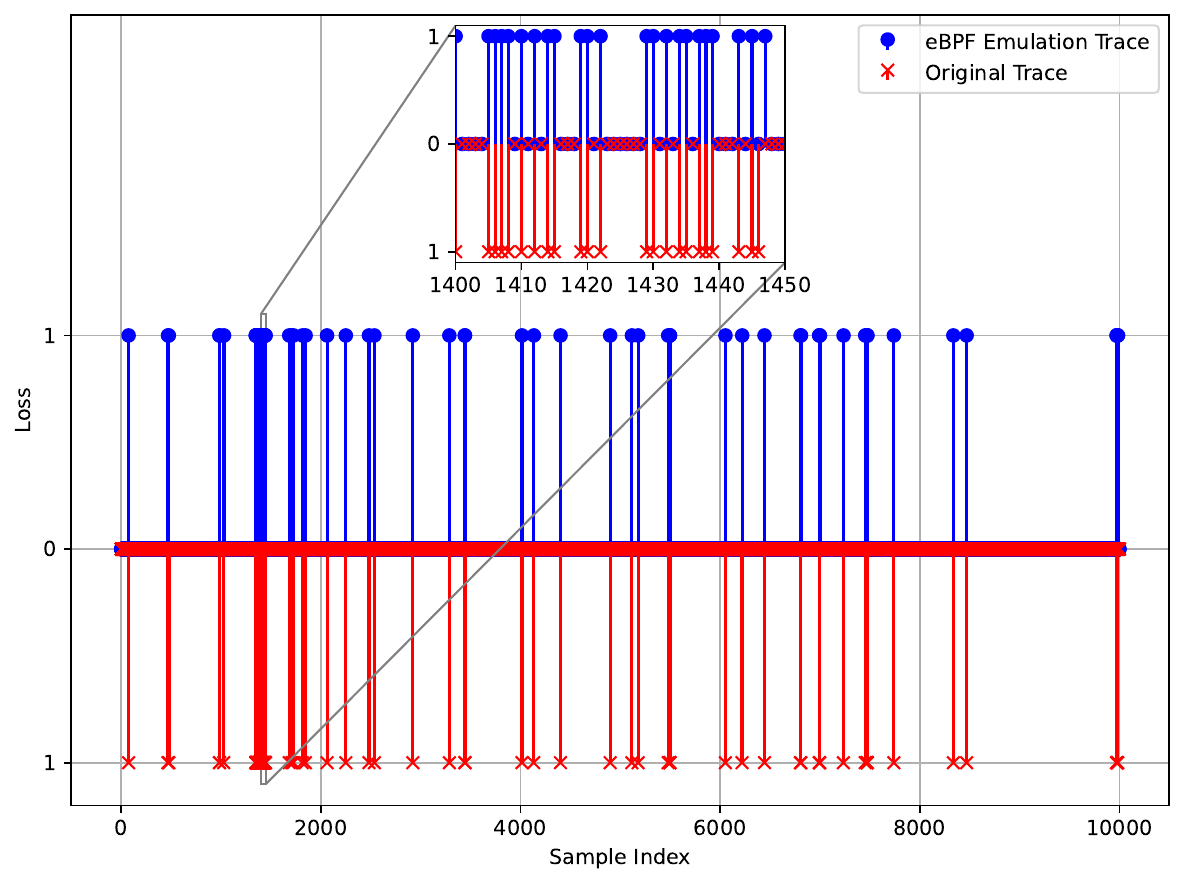}
  \caption{Comparison of the emulated and original loss traces}\label{fig_loss_comparison}
\end{figure}

Figs. \ref{fig_delay_comparison} and \ref{fig_loss_comparison} show $10000$ sample points of delay/loss traces of the original Starlink and the emulated connections, respectively. Fig. \ref{fig_delay_comparison} shows that the Starlink connection exhibits dynamic delay, where the minimum delay changes with a period of about $1500$ samples, corresponding to $15$s given $10$ms \texttt{irtt} sending interval. This has been confirmed in several Starlink measurement campaigns, where the change (and the abrupt jump) is due to handover of satellites. These characteristic are difficult to emulate using the  statistical models provided by the existing emulators. The figures show that the trace of the eBPF-emulated connection matches the original trace closely. From the zoomed-in subfigure, the emulated delay is only slightly higher (no more than $2$ms). This not only verifies that the emulation method is effective. Given that the packet sending is real, this also implies that the processing overhead/delay of the emulation is light, meaning that the method is very efficient. 


\section{Discussion and Conclusion}
This paper designed and implemented an eBPF-based trace-driven emulation method for satellite networks. Using traces collected from a Starlink testbed, we have verified that the proposed method can effectively and efficiently emulate the connection conditions, providing solid foundation for performance evaluation over such networks.

We point out that the proposed method is in fact a generic emulation approach. It can be a supplement/enhancement in many other scenarios than the satellite networks where the existing simulation or emulation methods cannot fulfill the requirement. For example, we may generate traces using protocol-based simulators and then use the proposed method to evaluation applications. This would be much faster compared to solely using the simulator, which introduces much overhead or has performance burden. 

\bibliographystyle{IEEEtran}

\begin{thebibliography}{1}
\providecommand{\url}[1]{#1}
\csname url@samestyle\endcsname
\providecommand{\newblock}{\relax}
\providecommand{\bibinfo}[2]{#2}
\providecommand{\BIBentrySTDinterwordspacing}{\spaceskip=0pt\relax}
\providecommand{\BIBentryALTinterwordstretchfactor}{4}
\providecommand{\BIBentryALTinterwordspacing}{\spaceskip=\fontdimen2\font plus
\BIBentryALTinterwordstretchfactor\fontdimen3\font minus \fontdimen4\font\relax}
\providecommand{\BIBforeignlanguage}[2]{{%
\expandafter\ifx\csname l@#1\endcsname\relax
\typeout{** WARNING: IEEEtran.bst: No hyphenation pattern has been}%
\typeout{** loaded for the language `#1'. Using the pattern for}%
\typeout{** the default language instead.}%
\else
\language=\csname l@#1\endcsname
\fi
#2}}
\providecommand{\BIBdecl}{\relax}
\BIBdecl

\bibitem{Puttonen2014SimuTools}
J.~Puttonen et al., ``Satellite model for network simulator 3.'' in \emph{SimuTools}, 2014, pp. 86--91.

\bibitem{Kassing2020IMC}
S.~Kassing et al., ``{Exploring the Internet from space with Hypatia},'' in \emph{Proc. ACM Internet Measurement Conf.}, 2020, pp. 214--229.

\bibitem{Kempton2021ICC}
B.~Kempton and A.~Riedl, ``Network simulator for large low earth orbit satellite networks,'' in \emph{IEEE Int. Conf. Commun. (ICC)}, 2021, pp. 1--6.

\bibitem{Ma2023Infocom}
S.~Ma et al., ``Network characteristics of {LEO} satellite constellations: A starlink-based measurement from end users,'' in \emph{IEEE INFOCOM}, 2023, pp. 1--10.

\bibitem{Pan2024ICC}
J.~Pan, J.~Zhao, and L.~Cai, ``Measuring the satellite links of a {LEO} network,'' in \emph{IEEE Int. Conf. Commun. (ICC)}, 2024, pp. 4439--4444.

\bibitem{Vieira2020ACMSurvey}
M.~A.~M. Vieira et al., ``Fast packet processing with {eBPF} and {XDP}: Concepts, code, challenges, and applications,'' \emph{ACM Comput. Surv.}, vol.~53, no.~1, Feb. 2020.

\bibitem{Saeed2017Sigcomm}
A.~Saeed et al., ``Carousel: Scalable traffic shaping at end hosts,'' in \emph{ACM SIGCOMM '17}, 2017, pp. 404--417.

\bibitem{Becker2022IC2E}
S.~Becker et al., ``Network emulation in large-scale virtual edge testbeds: A note of caution and the way forward,'' in \emph{IEEE Int. Conf. Cloud Engineering (IC2E)}, 2022, pp. 1--7.

\bibitem{irtt}
\BIBentryALTinterwordspacing
P.~Heist, ``Isochronous round-trip tester.'' [Online]. Available: \url{https://github.com/heistp/irtt}
\BIBentrySTDinterwordspacing

\end{thebibliography}

\end{document}